\documentclass[aps,preprint]{revtex4}%
\usepackage{amssymb}
\usepackage{amsfonts}
\usepackage{amsmath}
\usepackage{graphicx}%
\providecommand{\U}[1]{\protect\rule{.1in}{.1in}}

\begin{document}
\preprint{ }
\title[spontaneous emission]{Field energy and angular momentum in spontaneous emission: A
Schr\"{o}dinger-picture approach}
\author{P. R. Berman$^{1}$, A. Kuzmich$^{1}$, and P. W Milonni$^{2}$}
\affiliation{$^{1}$Physics Department, University of Michigan, Ann Arbor, MI \ 48109-1040}
\affiliation{$^{2}$Department of Physics and Astronomy, University of Rochester, Rochester,
New York 14627 USA}
\pacs{PACS number}

\begin{abstract}
A Schr\"{o}dinger-picture approach is used to calculate the field energy and
angular momentum radiated by an atom undergoing spontaneous emission. The
calculation is carried out using both the rotating-wave approximation (RWA)
and Weisskopf-Wigner approximation (WWA). It is shown that a consistent
application of the WWA leads to expressions for both the energy and angular
momentum in the field that are finite for all times, in contrast to the
results for a classical point dipole oscillator. Moreover, it is shown that
the total angular momentum in the field is a sum of its spin and orbital
components, again in contrast to the analogous results for a classical point
dipole oscillator. Analytic expressions for the energy, spin angular momentum,
and orbital angular momentum in the field are obtained for an atom undergoing
spontaneous emission from a state having angular momentum $H$ to a state
having angular momentum $G$ via an electric dipole transition. It is shown
that the spin and orbital angular momenta in the field are equal, independent
of the values of $H$ and $G$. With a slight modification of the WWA, it can
also be shown that the energy density in the field, the Poynting vector of the
field, and the angular momentum flux of the field do not diverge at the origin.

\end{abstract}
\volumeyear{year}
\volumenumber{number}
\issuenumber{number}
\eid{identifier}
\date[Date text]{date}
\received[Received text]{date}

\revised[Revised text]{date}

\accepted[Accepted text]{date}

\published[Published text]{date}

\startpage{1}
\maketitle

\section{Introduction}

In 1930, Weisskoff and Wigner addressed the question of spontaneous emission
by atoms \cite{ww}. To do so, they made a somewhat unphysical approximation,
assuming that the spectral density of the vacuum field was constant, extending
even to negative frequencies. With this approximation, they showed that the
excited state of the atom decayed exponentially and that the frequency
distribution of the radiated field was a Lorentzian (albeit extending to
negative frequencies). The Weisskopf-Wigner Approximation (WWA) leads to
results that are in excellent agreement with experiment, even if the
underlying assumptions used in its derivation are suspect. In fact, as far as
we know, there has been no experimental deviation from exponential decay for
atoms in free space. The dimensionless parameter that characterizes the
success of the WWA is the ratio of the decay rate to the transition frequency.
When this ratio is much less than unity (as it is for atomic transitions), the
success of the WWA is ensured.

The purpose of this paper is to show that, when applied consistently to
spontaneous emission, the WWA also leads to relatively simple expressions for
the energy and angular momentum of the radiated fields. In fact, analytic
expressions for both the spin and orbital angular components of the field are
obtained for atomic transitions between two states having arbitrary angular
momenta. Independent of the angular momentum of the atomic states, it is found
that the spin and orbital angular momentum components are equal to one
another. Moreover, in contrast to a source-field approach in which the energy
density of the field and the angular momentum flux diverge at the origin, we
find that the WWA produces energy densities and angular momentum flux that are
well-behaved at the origin. In addition, the angular momentum of the field is
a sum of its spin and orbital components, in contrast to what one would find
using a source-field approach \cite{bermcl}. We use a Schr\"{o}dinger-picture
approach to calculate the energy and angular momentum in the field. There have
been other calculations of the fields radiated in spontaneous emission
\cite{frbar,andr,barn2}, but the approaches taken in those works differ
considerably from the one taken in this paper. We consider energy as well as
angular momentum and aspects of WWA and RWA, neither of which are addressed in
Refs. \cite{frbar}-\cite{barn2}. Reference \cite{frbar} deals only with a
$J=1$ to $J=0$ transition and Refs. \cite{andr} and \cite{barn2} focus on
aspects of multipolar radiation fields. In Appendix B, we outline a WWA
calculation of the angular momentum in the field radiated spontaneously by a
charge in a uniform magnetic field.

\section{Basic Equations}

The atom is assumed to have a degenerate ground state with angular momentum
$G$ and a degenerate excited state with angular momentum $H$. The ground and
excited states are separated by frequency $\omega_{0}$ and the ground state
energy is set equal to zero. The Hamiltonian in the rotating-wave and dipole
approximation is given by%
\begin{equation}
H=H_{atom}+H_{f}+V,\label{ham1}%
\end{equation}
where
\begin{equation}
H_{atom}=\hbar\omega_{0}\sum_{m_{H}}\sigma_{Hm_{H};Hm_{H}}\label{hatom}%
\end{equation}
is the atomic Hamiltonian,%
\begin{equation}
H_{f}=\hbar\int d\mathbf{k}\sum_{\lambda}\omega_{k}a^{\dag}\left(
\mathbf{k}_{\lambda}\right)  a\left(  \mathbf{k}_{\lambda}\right)  \label{hf}%
\end{equation}
is the field Hamiltonian, and
\begin{equation}
V=-\frac{i}{\left(  2\pi\right)  ^{3/2}}\int d\mathbf{k}\left(  \frac
{\hbar\omega_{k}}{2\epsilon_{0}}\right)  ^{1/2}\sum_{\lambda}\left[
\boldsymbol{\mu}_{+}\cdot\boldsymbol{\epsilon}_{\mathbf{k}}^{(\lambda
)}a\left(  \mathbf{k}_{\lambda}\right)  -a^{\dag}\left(  \mathbf{k}_{\lambda
}\right)  \boldsymbol{\mu}_{-}\cdot\boldsymbol{\epsilon}_{\mathbf{k}%
}^{(\lambda)}\right]  \label{v}%
\end{equation}
is the atom-field interaction Hamiltonian. In these equations $\sigma
_{Hm_{H};Hm_{H}}$ is the atomic state population operator for the $m_{H}$
sublevel of level $H$, $a\left(  \mathbf{k}_{\lambda}\right)  $ and $a^{\dag
}\left(  \mathbf{k}_{\lambda}\right)  $ are annihilation and creation
operators for a field mode having wave vector $\mathbf{k}$, frequency
$\omega_{k}=kc$ and polarization $\lambda$ satisfying the commutation
relation
\begin{equation}
\left[  a\left(  \mathbf{k}_{\lambda}\right)  ,a^{\dag}\left(  \mathbf{k}%
_{\lambda^{\prime}}^{\prime}\right)  \right]  =\delta\left(  \mathbf{k}%
-\mathbf{k}^{\prime}\right)  \delta_{\lambda,\lambda^{\prime}},\label{comm}%
\end{equation}%
\begin{subequations}
\label{ep}%
\begin{align}
\boldsymbol{\epsilon}_{\mathbf{k}}^{(1)} &  =\boldsymbol{\epsilon}%
_{\mathbf{k}}^{(\theta)}=\cos\theta_{k}\cos\phi_{k}\mathbf{\hat{x}+}\cos
\theta_{k}\sin\phi_{k}\mathbf{\hat{y}-}\sin\theta_{k}\mathbf{\hat{z}%
,}\label{epa}\\
\boldsymbol{\epsilon}_{\mathbf{k}}^{(2)} &  =\boldsymbol{\epsilon}%
_{\mathbf{k}}^{(\phi)}=-\sin\phi_{k}\mathbf{\hat{x}+}\cos\phi_{k}%
\mathbf{\hat{y}}\label{epb}%
\end{align}
are the polarizations of mode $\mathbf{k}$,%
\end{subequations}
\begin{subequations}
\label{mu}%
\begin{align}
\boldsymbol{\mu}_{+}  & =\sum_{m_{G},m_{H}}\sum_{i=1}^{3}\left\langle
Hm_{H}\right\vert \mu_{i}\left\vert Gm_{G}\right\rangle \sigma_{Hm_{H};Gm_{G}%
}\mathbf{\hat{u}}_{i}\label{muplus}\\
\boldsymbol{\mu}_{-}  & =\sum_{m_{G},m_{H}}\sum_{i=1}^{3}\left\langle
Gm_{G}\right\vert \mu_{i}\left\vert Hm_{H}\right\rangle \sigma_{Gm_{G};Hm_{H}%
}\mathbf{\hat{u}}_{i}\label{muminus}%
\end{align}
are dipole moment operators, $\sigma_{Hm_{H};Gm_{G}}$ is a raising operator
from state $\left\vert Gm_{G}\right\rangle \ $to state $\left\vert
Hm_{H}\right\rangle $, $\sigma_{Gm_{G};Hm_{H}}$ is a lowering operator from
state $\left\vert Hm_{H}\right\rangle \ $to state $\left\vert Gm_{G}%
\right\rangle $, $\mu_{i}$ is the $i$th component of the dipole moment
operator, $\mathbf{\hat{u}}_{1}=\mathbf{\hat{x}}$, $\mathbf{\hat{u}}%
_{2}=\mathbf{\hat{y}}$, $\mathbf{\hat{u}}_{3}=\mathbf{\hat{z}}$\textbf{,
}$\delta\left(  \mathbf{k}-\mathbf{k}^{\prime}\right)  $ is a Dirac delta
function, $\delta_{\lambda,\lambda^{\prime}}$ is a Kronecker delta, and%
\end{subequations}
\begin{equation}
\int d\mathbf{k}=\int_{0}^{\infty}d\omega_{k}\frac{\omega_{k}^{2}}{c^{3}}\int
d\Omega_{k}.\label{wwap}%
\end{equation}
A caret over a vector indicates a unit vector. In writing Eq. (\ref{ham1}), we
have not included a term $\int\mathbf{P}^{2}d\mathbf{R}/\left(  2\epsilon
_{0}\right)  $, where $\mathbf{P}$ is the dipole moment density. This term
contributes to radiative energy shifts, but is not relevant to our discussion.
We have also neglected the (infinite) $\frac{1}{2}\sum_{\mathbf{k},\lambda
}\hbar\omega_{k}$ term corresponding to the vacuum field energy.

We have not yet made the WWA. To do so, in any subsequent final expressions
for the field energy or field angular momentum of the form $\int_{0}^{\infty
}d\omega_{k}\omega_{k}^{3}f\left(  \omega_{k}\right)  $, where $f\left(
\omega_{k}\right)  $ is sharply peaked at $\omega_{k}=\omega_{0}$, we set%
\begin{equation}
\int_{0}^{\infty}d\omega_{k}\omega_{k}^{3}f\left(  \omega_{k}\right)
\approx\omega_{0}^{3}\int_{-\infty}^{\infty}d\omega_{k}f\left(  \omega
_{k}\right)  . \label{wweq}%
\end{equation}
In effect this constitutes the WWA for spontaneous emission, corresponding to
a spectral density function for the vacuum field that is a constant for all frequencies.

Given an initial condition in which the atom is prepared in a linear
superposition of its excited state sublevels and the field is in the vacuum
state, in RWA we can write the state vector for times $t>0$ in an interaction
representation as
\begin{equation}
\left\vert \psi(t)\right\rangle =\sum_{m_{H}}c_{m_{H}}(t)e^{-i\omega_{0}%
t}\left\vert m_{H};0\right\rangle +\sum_{m_{G}}\int d\mathbf{k}\sum_{\lambda
}c_{m_{G}}^{(\lambda)}(\mathbf{k},t)e^{-i\omega_{k}t}\left\vert m_{G}%
;\mathbf{k}_{\lambda}\right\rangle ,\label{stvec}%
\end{equation}
where
\begin{equation}
\left\langle m_{G};\mathbf{k}_{\lambda}\right\vert \left.  m_{G}^{\prime
};\mathbf{k}_{\lambda^{\prime}}^{\prime}\right\rangle =\delta_{\lambda
,\lambda^{\prime}}\delta\left(  \mathbf{k}-\mathbf{k}^{\prime}\right)
\delta_{m_{G},m_{G}^{\prime}}.\label{orthog}%
\end{equation}
In these equations, $c_{m_{H}}(t)$ is the state amplitude for the atom to be
in state $\left\vert Hm_{H}\right\rangle $ and the field to be in the vacuum
state at time $t$ and $c_{m_{G}}^{(\lambda)}(\mathbf{k},t)$ is the state
amplitude for the atom to be in state $\left\vert Gm_{G}\right\rangle $ and
the field to be in state $\left\vert \mathbf{k}_{\lambda}\right\rangle $ at
time $t$. The state vector is a superposition of excited-state kets
$\left\vert m_{H};0\right\rangle $ corresponding to the atom in sublevel
$m_{H}$ and no photons in the field, and ground-state kets $\left\vert
m_{G};\mathbf{k}_{\lambda}\right\rangle $ corresponding to the atom in
sublevel $m_{G}$ with a photon having propagation vector $\mathbf{k}$ and
polarization $\lambda$ in the field.  The initial condition is%
\begin{equation}
\left\vert \psi(0)\right\rangle =\sum_{m_{H}}c_{m_{H}}(0)\left\vert
m_{H};0\right\rangle ,\label{psi0}%
\end{equation}
with%
\begin{equation}
\sum_{m_{H}}\left\vert c_{m_{H}}(0)\right\vert ^{2}=1.\label{init}%
\end{equation}
In the WWA, the excited-state amplitudes undergo exponential decay given by
\begin{equation}
c_{m_{H}}(t)=c_{m_{H}}(0)e^{-\gamma t},\label{expdec}%
\end{equation}
where the decay rate $\gamma$ is defined by
\begin{equation}
\gamma=\frac{\gamma_{H}}{2}=\frac{\omega_{0}^{3}\left\vert \mu_{HG}\right\vert
^{2}}{6\left(  2H+1\right)  \pi\epsilon_{0}\hbar c^{3}},\label{gam}%
\end{equation}
$\gamma_{H}$ is the decay rate of excited-state populations, and $\mu_{HG}$ is
a reduced matrix element. The time evolution equation for $c_{m_{G}}%
^{(\lambda)}(\mathbf{k},t)$ is
\begin{equation}
\dot{c}_{m_{G}}^{(\lambda)}(\mathbf{k},t)=\frac{1}{i\hbar}\sum_{m_{H}%
}\left\langle m_{G};\mathbf{k}_{\lambda}\right\vert V\left\vert m_{H}%
;0\right\rangle e^{i\left(  \omega_{k}-\omega_{0}\right)  t}c_{m_{H}%
}(0)e^{-\gamma t},
\end{equation}
with solution%
\begin{align}
c_{m_{G}}^{(\lambda)}(\mathbf{k},t) &  =\frac{\mu_{HG}^{\ast}}{\left(
2\pi\right)  ^{3/2}}\left(  \frac{\omega_{k}}{2\hbar\epsilon_{0}}\right)
^{1/2}e^{-i\left(  \omega_{0}-\omega_{k}\right)  t}\sum_{m_{H}}\tilde
{V}_{m_{G},m_{H}}^{\left(  \lambda\right)  }\left(  \Omega_{k}\right)
c_{m_{H}}(0)\nonumber\\
&  \times\frac{e^{-i\left(  \omega_{k}-\omega_{0}\right)  t}-e^{-\gamma t}%
}{\gamma-i\left(  \omega_{k}-\omega_{0}\right)  },\label{examp}%
\end{align}
where
\begin{equation}
\tilde{V}_{m_{G},m_{H}}^{\left(  \lambda\right)  }\left(  \Omega_{k}\right)
=\frac{1}{\left(  2G+1\right)  ^{1/2}}\sum_{q=-1}^{1}%
\begin{bmatrix}
H & 1 & G\\
m_{H} & -q & m_{G}%
\end{bmatrix}
(-1)^{q}\epsilon_{q}^{\left(  \lambda\right)  }\left(  \Omega_{k}\right)  ,
\end{equation}
the quantity in square brackets is a Clebsch-Gordan coefficient, and
\begin{subequations}
\label{pol}%
\begin{align}
\epsilon_{q}^{\left(  \theta\right)  }\left(  \Omega_{k}\right)   &
=-\sin\theta_{k}\delta_{q,0}-\frac{\cos\theta_{k}}{\sqrt{2}}\left(
e^{i\phi_{k}}\delta_{q,1}-e^{-i\phi_{k}}\delta_{q,-1}\right)  ,\label{pola}\\
\epsilon_{q}^{\left(  \phi\right)  }\left(  \Omega_{k}\right)   &
=0\delta_{q,0}-\frac{i}{\sqrt{2}}\left(  e^{i\phi_{k}}\delta_{q,1}%
+e^{-i\phi_{k}}\delta_{q,-1}\right)  ,\label{polb}%
\end{align}
where $\delta_{q,q^{\prime}}$ is a Kronecker delta.

\section{Energy}

It is clear that energy must be conserved for the time-independent Hermitian
Hamiltonian given in Eq. (\ref{ham1}). However this does not necessarily imply
that the energy of the field plus the atom is conserved. In fact it was shown
that, if the WWA is not made, different spectral densities can lead to a field
energy that is many times $\hbar\omega_{0}$ for times less than the decay time
\cite{bf}. At such times, the interaction energy is negative ensuring that the
sum of field energy plus atom energy plus interaction energy is equal to
$\hbar\omega_{0}$. Moreover, even though energy is conserved for the
Hamiltonian given in Eq. (\ref{ham1}), we are not guaranteed that energy will
be conserved when the WWA given in Eq. (\ref{wweq}) is used.

In general, for the state vector given in Eq. (\ref{stvec}) it is not
difficult to show that
\end{subequations}
\begin{subequations}
\label{dd}%
\begin{align}
\left\langle H_{atom}(t)\right\rangle  &  =\hbar\omega_{0}\sum_{m_{H}%
}\left\vert c_{m_{H}}(t)\right\vert ^{2}=\hbar\omega_{0}e^{-\gamma_{H}%
t},\label{dda}\\
\left\langle H_{f}(t)\right\rangle  &  =\hbar\sum_{m_{G}}\int d\mathbf{k}%
\omega_{k}\sum_{\lambda}\left\vert c_{m_{G}}^{(\lambda)}(\mathbf{k}%
,t)\right\vert ^{2}.\label{ddb}%
\end{align}
Using Eqs. (\ref{examp}), (\ref{wwap}), (\ref{wweq}), (\ref{expdec}), and
(\ref{gam}), and carrying out the various summations and integrals in Eq.
(\ref{ddb}), we find that%
\end{subequations}
\begin{align}
\left\langle H_{f}(t)\right\rangle  &  =\frac{\hbar\gamma}{\pi}\int_{-\infty
}^{\infty}d\omega_{k}\frac{1-e^{-\gamma t}-2\cos\left[  \left(  \omega
_{k}-\omega_{0}\right)  t\right]  e^{-\gamma t}}{\gamma^{2}+\left(  \omega
_{k}-\omega_{0}\right)  ^{2}}\omega_{k}\nonumber\\
&  =\hbar\omega_{0}\left(  1-e^{-\gamma_{H}t}\right)  ,\label{feg}%
\end{align}
such that
\begin{equation}
\left\langle H_{atom}(t)\right\rangle +\left\langle H_{f}(t)\right\rangle
=\hbar\omega_{0};
\end{equation}
the sum of atomic and field energies is conserved. Moreover, it is possible to
use Eqs. (\ref{v}), (\ref{examp}), and (\ref{wweq}) to show explicitly that
$\left\langle V(t)\right\rangle =0$. Alternatively, using Eqs. (\ref{ham1}%
)-(\ref{v}) and the fact that probability is conserved, one finds that
\begin{equation}
\left\langle V(t)\right\rangle =\hbar\sum_{\lambda}\int d\mathbf{k}\left(
\omega_{k}-\omega_{0}\right)  \left\vert c_{m_{G}}^{(\lambda)}(\mathbf{k}%
,t)\right\vert ^{2}.
\end{equation}
In the WWA, $\left\langle V(t)\right\rangle =0$ since $\left\vert c_{m_{G}%
}^{(\lambda)}(\mathbf{k},t)\right\vert ^{2}$ is an even function $\left(
\omega_{k}-\omega_{0}\right)  $ and the integral over $\omega_{k}$ is from
$-\infty$ to $\infty$. As a consequence, the average total energy,
$\left\langle H_{atom}(t)\right\rangle +\left\langle H_{f}(t)\right\rangle
+\left\langle V(t)\right\rangle $, is also conserved, although this wasn't
guaranteed using our prescription for the WWA.

\section{Angular Momentum}

We shall assume that the quantized field operators are given by
\begin{equation}
\mathbf{E}(\mathbf{R})=\frac{i}{\left(  2\pi\right)  ^{3/2}}\int
d\mathbf{k}\left(  \frac{\hbar\omega_{k}}{2\epsilon_{0}}\right)  ^{1/2}%
\sum_{\lambda}\left[  a\left(  \mathbf{k}_{\lambda}\right)  e^{i\mathbf{k}%
\cdot\mathbf{R}}-a^{\dag}\left(  \mathbf{k}_{\lambda}\right)  e^{-i\mathbf{k}%
\cdot\mathbf{R}}\right]  \boldsymbol{\epsilon}_{\mathbf{k}}^{(\lambda
)},\label{ef}%
\end{equation}%
\begin{equation}
\mathbf{B}(\mathbf{R})=\frac{i}{\left(  2\pi\right)  ^{3/2}c}\int
d\mathbf{k}\left(  \frac{\hbar\omega_{k}}{2\epsilon_{0}}\right)  ^{1/2}%
\sum_{\lambda}\left[  a\left(  \mathbf{k}_{\lambda}\right)  e^{i\mathbf{k}%
\cdot\mathbf{R}}-a^{\dag}\left(  \mathbf{k}_{\lambda}\right)  e^{-i\mathbf{k}%
\cdot\mathbf{R}}\right]  \mathbf{\hat{k}\times}\boldsymbol{\epsilon
}_{\mathbf{k}}^{(\lambda)},\label{bf}%
\end{equation}
and that the vector potential operator in Coulomb gauge by%
\begin{equation}
\mathbf{A}(\mathbf{R})=\frac{1}{\left(  2\pi\right)  ^{3/2}}\int
d\mathbf{k}\left(  \frac{\hbar}{2\epsilon_{0}\omega_{k}}\right)  ^{1/2}%
\sum_{\lambda}\left[  a\left(  \mathbf{k}_{\lambda}\right)  e^{i\mathbf{k}%
\cdot\mathbf{R}}+a^{\dag}\left(  \mathbf{k}_{\lambda}\right)  e^{-i\mathbf{k}%
\cdot\mathbf{R}}\right]  \boldsymbol{\epsilon}_{\mathbf{k}}^{(\lambda
)}.\label{af}%
\end{equation}
The total angular momentum operator of the field is defined by%
\begin{equation}
\mathbf{L}_{f}=\epsilon_{0}\int d\mathbf{R\,R}\times\left[  \mathbf{E}%
(\mathbf{R})\times\mathbf{B}(\mathbf{R})\right]  .
\end{equation}
Often, vector identities are used to decompose the angular momentum operators
into "spin" and an "orbital" parts, given by \cite{simm,mc3}
\begin{subequations}
\label{angdecomp}%
\begin{align}
\mathbf{L}_{spin} &  =\epsilon_{0}\int d\mathbf{R\,E}(\mathbf{R}%
)\times\mathbf{A}(\mathbf{R}),\label{angspin}\\
\mathbf{L}_{orb} &  =\epsilon_{0}\int d\mathbf{R\,}\sum_{j=1}^{3}%
E_{j}(\mathbf{R})\left(  \mathbf{R}\times\mathbf{\nabla}\right)
A_{j}(\mathbf{R}),\label{angorb}%
\end{align}
where $E_{j}(\mathbf{R})$ and $A_{j}(\mathbf{R})$ are the $j$th component of
the electric field and vector potential operators, respectively.

It will prove useful to use Eqs. (\ref{ef})-(\ref{angdecomp}) to express the
angular momentum operator in terms of field creation and annihilation
operators. In Appendix A it is shown that, with the neglect of terms varying
as $a_{\mathbf{k}_{\lambda}}a_{\mathbf{k}_{\lambda^{\prime}}^{\prime}}$ or
$a_{\mathbf{k}_{\lambda}}^{\dag}a_{\mathbf{k}_{\lambda^{\prime}}^{\prime}%
}^{\dag}$ which vanish when taking expectation values, the total angular
momentum operator of the field is given by
\end{subequations}
\begin{align}
\mathbf{L}_{f}  &  =\left(  i\frac{\hbar}{2c}\sum_{\lambda,\lambda^{\prime}%
}\int d\mathbf{k\,}\sqrt{\omega_{k}}a^{\dag}\mathbf{\left(  \mathbf{k}%
_{\lambda^{\prime}}\right)  }\left[  \left(  \mathbf{\hat{k}\times
}\boldsymbol{\epsilon}_{\mathbf{k}}^{(\lambda^{\prime})}\right)
\cdot\mathbf{\nabla}_{\mathbf{k}}\right]  \left[  \sqrt{\omega_{k}}a\left(
\mathbf{k}_{\lambda}\right)  \boldsymbol{\epsilon}_{\mathbf{k}}^{(\lambda
)}\right]  +adj\right) \nonumber\\
&  +\left(  -i\frac{\hbar}{2c}\sum_{\lambda,\lambda^{\prime}}\int
d\mathbf{k\,}\sqrt{\omega_{k}}a^{\dag}\mathbf{\left(  \mathbf{k}%
_{\lambda^{\prime}}\right)  }\left(  \mathbf{\hat{k}\times}%
\boldsymbol{\epsilon}_{\mathbf{k}}^{(\lambda^{\prime})}\right)  \mathbf{\nabla
}_{\mathbf{k}}\cdot\left[  \sqrt{\omega_{k}}a\left(  \mathbf{k}_{\lambda
}\right)  \boldsymbol{\epsilon}_{\mathbf{k}}^{(\lambda)}\right]  +adj\right)
, \label{eltot}%
\end{align}
where "$adj$" stands for "adjoint." The spin angular and orbital angular
momenta field operators can be written as \cite{cc,enk,bw,barn3}%
\begin{equation}
\mathbf{L}_{spin}=\frac{i\hbar}{\left(  2\pi\right)  ^{3}}\int d\mathbf{k\,}%
a\left(  \mathbf{k}_{\theta}\right)  a^{\dag}\left(  \mathbf{k}_{\phi}\right)
\mathbf{\hat{k}}+adj \label{elspin}%
\end{equation}
and%
\begin{equation}
\mathbf{L}_{orb}=\frac{i\hbar}{2}\sum_{j=1}^{3}\int d\mathbf{k}\sum
_{\lambda,\lambda^{\prime}}a\left(  \mathbf{k}_{\lambda^{\prime}}\right)
\left(  \boldsymbol{\epsilon}_{\mathbf{k}}^{(\lambda^{\prime})}\right)
_{j}\left(  -\mathbf{k\times\nabla}_{\mathbf{k}}\right)  \left[  a^{\dag
}\left(  \mathbf{k}_{\lambda}\right)  \left(  \boldsymbol{\epsilon
}_{\mathbf{k}}^{(\lambda)}\right)  _{j}\right]  +adj. \label{elorb}%
\end{equation}
It is not necessarily true that $\left\langle \mathbf{L}(t)\right\rangle
=\left\langle \mathbf{L}_{spin}(t)\right\rangle +\left\langle \mathbf{L}%
_{orb}(t)\right\rangle $ since there is also a surface term that can
contribute \cite{bermcl}.

It is now a simple matter to calculate the expectation values of the various
angular momenta operators. We will calculate only the $z-$component, but it is
not difficult to calculate the other components as well. For the atom, some of
the angular momentum is transferred to the ground state as the atom undergoes
spontaneous emission. Explicitly,
\begin{equation}
\left\langle L_{atom}(t)\right\rangle _{z}=\hbar e^{-\gamma_{H}t}\sum_{m_{H}%
}\left\vert c_{m_{H}}(0)\right\vert ^{2}m_{H}+\hbar\sum_{m_{G}}\sum_{\lambda
}\int d\mathbf{k}\left\vert c_{m_{G}}^{(\lambda)}(\mathbf{k},t)\right\vert
^{2}m_{G}.
\end{equation}
Using Eqs. (\ref{examp}), (\ref{wwap}), (\ref{expdec}), and (\ref{wweq}), and
carrying out the various summations and integrals, we obtain
\begin{equation}
\left\langle L_{atom}(t)\right\rangle _{z}=\hbar e^{-\gamma_{H}t}\sum_{m_{H}%
}\left\vert c_{m_{H}}(0)\right\vert ^{2}m_{H}+\hbar\left(  1-e^{-\gamma_{H}%
t}\right)  \sum_{m_{G},m_{H}}m_{G}%
\begin{bmatrix}
G & 1 & H\\
m_{G} & m_{H}-m_{G} & m_{H}%
\end{bmatrix}
^{2}\left\vert c_{m_{H}}(0)\right\vert ^{2}.
\end{equation}
The sum over $m_{G}$ can be carried out, resulting in%
\begin{align}
\left\langle L_{atom}(t)\right\rangle _{z}  &  =\hbar\left\{  e^{-\gamma_{H}%
t}+\left(  1-e^{-\gamma_{H}t}\right)  \left[  \frac{H-1}{H}\delta
_{H,G+1}+\frac{H\left(  H+1\right)  -1}{H\left(  H+1\right)  }\delta
_{H,G}+\frac{H+2}{H+1}\delta_{H,G-1}\right]  \right\} \nonumber\\
&  \times\sum_{m_{H}}\left\vert c_{m_{H}}(0)\right\vert ^{2}m_{H}.
\label{elavg}%
\end{align}
For $G=0$ and $H=1$, the term in square brackets is equal to zero and
$\left\langle L_{atom}(\infty)\right\rangle _{z}=0$.

The expectation value of total field angular momentum can be calculated from
Eqs. (\ref{eltot}) and (\ref{stvec}) as%
\begin{align}
\left\langle \mathbf{L}_{f}(t)\right\rangle  &  =\left(  i\frac{\hbar}{2c}%
\sum_{\lambda,\lambda^{\prime}}\sum_{m_{G}}\int d\mathbf{k\,}\sqrt{\omega_{k}%
}c_{m_{G}}^{(\lambda^{\prime})}(\mathbf{k},t)^{\ast}\left[  \left(
\mathbf{\hat{k}\times}\boldsymbol{\epsilon}_{\mathbf{k}}^{(\lambda^{\prime}%
)}\right)  \cdot\mathbf{\nabla}_{\mathbf{k}}\right]  \left[  \sqrt{\omega_{k}%
}c_{m_{G}}^{(\lambda)}(\mathbf{k},t)\boldsymbol{\epsilon}_{\mathbf{k}%
}^{(\lambda)}\right]  +c.c.\right) \nonumber\\
&  +\left(  -i\frac{\hbar}{2c}\sum_{\lambda,\lambda^{\prime}}\sum_{m_{G}}\int
d\mathbf{k\,}\sqrt{\omega_{k}}c_{m_{G}}^{(\lambda^{\prime})}(\mathbf{k}%
,t)^{\ast}\left(  \mathbf{\hat{k}\times}\boldsymbol{\epsilon}_{\mathbf{k}%
}^{(\lambda^{\prime})}\right)  \mathbf{\nabla}_{\mathbf{k}}\cdot\left[
\sqrt{\omega_{k}}c_{m_{G}}^{(\lambda)}(\mathbf{k},t)\boldsymbol{\epsilon
}_{\mathbf{k}}^{(\lambda)}\right]  +c.c.\right)  .
\end{align}
where $c.c.$ stands for "complex conjugate." The first term vanishes owing to
the integral over solid angle (in $k-$space) and the $z-$component of the
second term reduces to%
\begin{equation}
\left\langle L_{f}(t)\right\rangle _{z}=-i\frac{\hbar}{2}\sum_{m_{G}}\int
d\mathbf{k\,}c_{m_{G}}^{(\phi)}(\mathbf{k},t)^{\ast}\left[  \frac{\partial
}{\partial\theta_{k}}\left(  \sin\theta_{k}c_{m_{G}}^{(\theta)}(\mathbf{k}%
,t)\right)  +\frac{\partial}{\partial\phi_{k}}\left(  c_{m_{G}}^{(\phi
)}(\mathbf{k},t)\right)  \right]  +c.c. \label{angtot}%
\end{equation}
Using Eq. (\ref{examp}) and the WWA prescription defined in Eq. (\ref{wweq}),
we find%
\begin{align}
\left\langle L_{f}(t)\right\rangle _{z}  &  =-i\frac{\hbar}{2}\frac{\left\vert
\mu_{HG}\right\vert ^{2}}{\left(  2\pi\right)  ^{3}}\left(  \frac{\omega
_{0}^{3}}{2\hbar\epsilon_{0}}\right)  \sum_{m_{G},m_{H},m_{H}^{\prime}}\left[
c_{m_{H}}(0)\right]  ^{\ast}c_{m_{H^{\prime}}}(0)\nonumber\\
&  \times\int_{-\infty}^{\infty}d\omega_{k}\frac{\left\vert e^{-i\left(
\omega_{k}-\omega_{0}\right)  t}-e^{-\gamma t}\right\vert ^{2}}{\gamma
^{2}+\left(  \omega_{k}-\omega_{0}\right)  ^{2}}\nonumber\\
&  \times\int d\Omega_{k}\mathbf{\,}\left[  \tilde{V}_{m_{G},m_{H}}^{\left(
\phi\right)  }\left(  \Omega_{k}\right)  \right]  ^{\ast}\left[
\frac{\partial}{\partial\theta_{k}}\left(  \sin\theta_{k}\tilde{V}%
_{m_{G},m_{H}^{\prime}}^{\left(  \theta\right)  }\left(  \Omega_{k}\right)
\right)  +\frac{\partial}{\partial\phi_{k}}\left(  \tilde{V}_{m_{G}%
,m_{H}^{\prime}}^{\left(  \phi\right)  }\left(  \Omega_{k}\right)  \right)
\right]  +c.c.
\end{align}
The integral over $\phi_{k}$ is proportional to $2\pi\delta_{m_{H}%
,m_{H^{\prime}}}$, where $\delta_{m,n}$ is a Kronecker delta. The sums and
integrals can be carried out to arrive at%
\begin{equation}
\left\langle L_{f}(t)\right\rangle _{z}=\hbar\left(  1-e^{-\gamma_{H}%
t}\right)  \sum_{m_{H}}m_{H}\left\vert c_{m_{H}}(0)\right\vert ^{2}\left[
\frac{1}{H}\delta_{H,G+1}+\frac{1}{H\left(  H+1\right)  (2H+1)}\delta
_{H,G}-\frac{1}{\left(  H+1\right)  }\delta_{H,G-1}\right]  ,
\end{equation}
where Eq. (\ref{gam}) was used.

In a similar fashion, the expectation values of the spin and orbital angular
momenta of the field can be calculated from Eqs. (\ref{elspin}),
(\ref{elorb}), and (\ref{stvec}) as
\begin{equation}
\left\langle L_{spin}(t)\right\rangle _{z}=\left\langle L_{orb}%
(t)\right\rangle _{z}=\left\langle L_{f}(t)\right\rangle _{z}/2.
\end{equation}
Thus, the angular momentum is the same for the spin and orbital parts and the
total angular momentum is conserved,
\begin{equation}
\left\langle L_{atom}(t)\right\rangle _{z}+\left\langle L_{spin}%
(t)\right\rangle _{z}+\left\langle L_{orb}(t)\right\rangle _{z}=\left\langle
L_{atom}(0)\right\rangle _{z}.
\end{equation}
For $G=0$ and $H=1$,
\begin{equation}
\left\langle L_{spin}(\infty\right\rangle _{z}=\left\langle L_{orb}%
(\infty)\right\rangle _{z}=\frac{\hbar}{2}\sum_{m_{H}}m_{H}\left\vert
c_{m_{H}}(0)\right\vert ^{2}=\frac{\left\langle L_{atom}(0)\right\rangle _{z}%
}{2}.
\end{equation}

\section{Potential Limitations of the Rotating-Wave and Weisskopf-Wigner
Approximations}

It is known that the rotating wave approximation (RWA) leads to acausal
behavior of the fields radiated by an atom \cite{mil}. That is, within the
RWA, quantities such as the expectation of the Poynting vector and energy
density of the radiation fields are not fully retarded. Moreover, we have
already noted that the WWA is unphysical since it extends to negative
frequencies. The question arises, however, as to whether or not two wrongs can
make a right. Can the use of both the RWA and WWA lead to field quantities
that are fully retarded? The answer to this is a qualified "yes," but there is
a price to pay.

So far, we have calculated the total energy and angular momentum in the field.
Questions of retardation do not enter in such calculations. It is only when we
calculate quantities such as the Poynting vector or energy density or angular
momentum flux density \cite{barn} that we encounter results that depend
critically on the retarded nature of the fields. To illustrate matters we will
calculate the radial energy density for the fields for an atom undergoing
spontaneous emission on a $H=1$ to $G=0$ transition. Similar considerations
apply for the Poynting vector and angular momentum flux. We will take as our
initial state vector,
\begin{equation}
\left\vert \psi(0)\right\rangle =\left\vert e;0\right\rangle ,
\end{equation}
where $e$ corresponds to the $m_{H}=0$ sublevel. With this initial condition
and $c_{m_{G}}^{(\lambda)}(\mathbf{k},t)\equiv c_{g}^{(\lambda)}%
(\mathbf{k},t)$, it follows from Eqs. (\ref{examp}) - (\ref{pol}) that
\begin{align}
c_{g}^{(\theta)}(\mathbf{k},t) &  =-\frac{\mu_{HG}^{\ast}}{\left(
2\pi\right)  ^{3/2}}\left(  \frac{\omega_{k}}{2\hbar\epsilon_{0}}\right)
^{1/2}\frac{\sin\theta_{k}}{\sqrt{3}}e^{i\left(  \omega_{k}-\omega_{0}\right)
t}\nonumber\\
&  \times\frac{e^{-i\left(  \omega_{k}-\omega_{0}\right)  t}-e^{-\gamma t}%
}{\gamma-i\left(  \omega_{k}-\omega_{0}\right)  }\label{apk}%
\end{align}
and $c_{g}^{(\phi)}(\mathbf{k},t)=0$.

The average radial energy density is defined by%
\begin{align}
\left\langle w(R,t)\right\rangle  &  =\epsilon_{0}R^{2}\int d\Omega
\left\langle \mathbf{E}_{-}(\mathbf{R})\cdot\mathbf{E}_{+}(\mathbf{R}%
)+c^{2}\mathbf{B}_{-}(\mathbf{R})\cdot\mathbf{B}_{+}(\mathbf{R})\right\rangle
\nonumber\\
&  =\left(  \frac{\hbar}{2}\right)  \frac{R^{2}}{\left(  2\pi\right)  ^{3}%
}\int d\Omega\int d\mathbf{k}\int d\mathbf{k}^{\prime}\sqrt{\omega_{k}%
\omega_{k^{\prime}}}\nonumber\\
&  \times\left\langle a^{\dag}\left(  \mathbf{k}_{\lambda^{\prime}}^{\prime
}\right)  a\left(  \mathbf{k}_{\lambda}\right)  \right\rangle e^{-i\mathbf{k}%
^{\prime}\cdot\mathbf{R}}e^{i\mathbf{k}\cdot\mathbf{R}}\left[
\boldsymbol{\epsilon}_{\mathbf{k}}^{(\lambda)}\cdot\boldsymbol{\epsilon
}_{\mathbf{k}^{\prime}}^{(\lambda^{\prime})}+\left(  \mathbf{\hat{k}}^{\prime
}\mathbf{\times}\boldsymbol{\epsilon}_{\mathbf{k}^{\prime}}^{(\lambda^{\prime
})}\right)  \cdot\left(  \mathbf{\hat{k}\times}\boldsymbol{\epsilon
}_{\mathbf{k}}^{(\lambda)}\right)  \right]  , \label{ggg}%
\end{align}
where $\mathbf{E}_{+}(\mathbf{R})=\left[  \mathbf{E}_{-}(\mathbf{R})\right]
^{\dag}$ is given by the first term in Eq. (\ref{ef}) and $\mathbf{B}%
_{+}(\mathbf{R})=\left[  \mathbf{B}_{-}(\mathbf{R})\right]  ^{\dag}$ by the
first term in Eq. (\ref{bf}). Equation (\ref{ggg}) can be rewritten as
\begin{align}
\left\langle w(R,t)\right\rangle  &  =\left(  \frac{\hbar}{2}\right)
\frac{R^{2}}{\left(  2\pi\right)  ^{3}}\int d\Omega\left\vert \int
d\mathbf{k}\sqrt{\omega_{k}}c_{g}^{(\theta)}(\mathbf{k},t)e^{i\mathbf{k}%
\cdot\mathbf{R}}\boldsymbol{\epsilon}_{\mathbf{k}}^{(\theta)}\right\vert
^{2}\nonumber\\
&  +\left(  \frac{\hbar}{2}\right)  \frac{R^{2}}{\left(  2\pi\right)  ^{3}%
}\int d\Omega\left\vert \int d\mathbf{k}\sqrt{\omega_{k}}c_{g}^{(\theta
)}(\mathbf{k},t)e^{i\mathbf{k}\cdot\mathbf{R}}\boldsymbol{\epsilon
}_{\mathbf{k}}^{(\phi)}\right\vert ^{2}. \label{wr}%
\end{align}
Substituting Eq. (\ref{apk}) into Eq. (\ref{wr}), expanding $e^{i\mathbf{k}%
\cdot\mathbf{R}}$ in terms of spherical harmonics, and carrying out the
angular integrations, we find%
\begin{align}
\left\langle w(R,t)\right\rangle  &  =\frac{\left\vert \mu_{HG}\right\vert
^{2}R^{2}}{54\epsilon_{0}\pi^{3}c^{6}}\left[  2\left\vert P_{0}\right\vert
^{2}+3\left\vert P_{1}\right\vert ^{2}+\left\vert P_{2}\right\vert ^{2}\right]
\nonumber\\
&  =\frac{\hbar\gamma R^{2}}{3\pi^{2}\omega_{0}^{3}c^{3}}\left[  2\left\vert
P_{0}\right\vert ^{2}+3\left\vert P_{1}\right\vert ^{2}+\left\vert
P_{2}\right\vert ^{2}\right]  , \label{wr1}%
\end{align}
where
\begin{equation}
P_{\alpha}(R,t)=\int_{0}^{\infty}d\omega_{k}\,\omega_{k}^{3}\frac{e^{-i\left(
\omega_{k}-\omega_{0}\right)  t}-e^{-\gamma t}}{\gamma-i\left(  \omega
_{k}-\omega_{0}\right)  }j_{\alpha}(kR), \label{pj}%
\end{equation}
and $j_{\alpha}(kR)$ is a spherical Bessel function.

Before analyzing this result, let us calculate the expectation value of the
total field energy,%
\begin{equation}
W_{f}(t)=\int_{0}^{\infty}dR\left\langle w(R,t)\right\rangle ,
\end{equation}
to see if energy is conserved. Using the identities
\begin{align}
\int_{0}^{\infty}dRR^{2}j_{0}(kR)j_{0}(k^{\prime}R) &  =\int_{0}^{\infty
}dRR^{2}j_{2}(kR)j_{2}(k^{\prime}R)=\frac{\pi}{2k^{2}}\left[  \delta\left(
k-k^{\prime}\right)  +\delta\left(  k+k^{\prime}\right)  \right]  ,\nonumber\\
\int_{0}^{\infty}dRR^{2}j_{1}(kR)j_{1}(k^{\prime}R) &  =\frac{\pi}{2k^{2}%
}\left[  \delta\left(  k-k^{\prime}\right)  -\delta\left(  k+k^{\prime
}\right)  \right]  ,
\end{align}
we find
\begin{equation}
W_{f}(t)=\frac{\hbar\gamma}{\pi\omega_{0}^{3}}\int_{0}^{\infty}d\omega
_{k}\,\omega_{k}^{4}\frac{\left\vert e^{-i\left(  \omega_{k}-\omega
_{0}\right)  t}-e^{-\gamma t}\right\vert ^{2}}{\gamma^{2}+\left(  \omega
_{k}-\omega_{0}\right)  ^{2}}.
\end{equation}
If we apply the prescription for using the WWA given in Eq. (\ref{wweq}), we
obtain%
\begin{equation}
W_{f}(t)=\frac{\hbar\gamma}{\pi}\int_{-\infty}^{\infty}d\omega_{k}\,\omega
_{k}\frac{\left\vert e^{-i\left(  \omega_{k}-\omega_{0}\right)  t}-e^{-\gamma
t}\right\vert ^{2}}{\gamma^{2}+\left(  \omega_{k}-\omega_{0}\right)  ^{2}%
}=\hbar\omega_{0}\left(  1-e^{-\gamma_{H}t}\right)  .
\end{equation}
consistent with Eq. (\ref{feg}).

We now return to Eqs. (\ref{wr1}) and (\ref{pj}). If we apply our prescription
(\ref{wweq}) to Eq. (\ref{pj}), we find an average energy density given by%
\begin{equation}
\left\langle w^{WW}(R,t)\right\rangle =\frac{\hbar\gamma R^{2}}{3\pi^{2}c^{3}%
}\left[  2\left\vert Q_{0}\right\vert ^{2}+3\left\vert Q_{1}\right\vert
^{2}+\left\vert Q_{2}\right\vert ^{2}\right]  ,\label{edww}%
\end{equation}
where
\begin{equation}
Q_{\alpha}(R,t)=\int_{-\infty}^{\infty}d\omega\,\frac{e^{-i\omega
t}-e^{-\gamma t}}{\gamma-i\omega}j_{\alpha}\left[  x\left(  1+\frac{\omega
}{\omega_{0}}\right)  \right]  ,\label{qww}%
\end{equation}
and
\begin{equation}
x=k_{0}R=\omega_{0}R/c.\label{ex}%
\end{equation}
In this limit, there are major contributions to the integrand in Eq.
(\ref{qww}) at $\omega=0$ and for $\omega=-\omega_{0}$. The average energy
density given in Eq. (\ref{edww}) is well-behaved in the limit as
$x=k_{0}R\rightarrow0$, in contrast to the energy density for a classical
dipole which diverges as $R\rightarrow0$. This is a satisfying result. In the
WWA, the expectation value of the interaction energy vanishes, so we should
not expect a divergent energy density. On the other hand, the contribution to
the integral in Eq. (\ref{qww}) at $\omega=-\omega_{0}$ poses some serious
problems. Worst of all is that it predicts an energy density that is
nonvanishing for $t<R/c$.

Typically, $\omega_{0}\gtrsim10^{6}\gamma$ for electric dipole transitions. To
avoid problems associated with the contribution to the integral in Eq.
(\ref{qww}) at $\omega=-\omega_{0}$, we impose a cutoff at $\omega
=-1000\gamma$ [that is, for $\omega_{kc}=\omega_{0}-1000\gamma$, where
$\omega_{kc}$ is the cutoff frequency in the sum over field modes] . We refer
to this as a modified WWA (MWWA). It will not affect the energy calculations
in any significant way. Moreover, it is consistent with a natural cutoff that
occurs when the atom is assumed to be in its ground state and is then excited
by a classical radiation pulse whose duration $T$ satisfies $\omega_{0}%
T\gg\left(  \omega_{kc}-\omega_{0}\right)  T\gg1\gg\gamma T$ \cite{berww}. In
the MWWA, we can replace $j_{\alpha}\left[  x\left(  1+\frac{\omega}%
{\omega_{0}}\right)  )\right]  $ by $j_{\alpha}(x)$ unless $x\gtrsim1000$.
Thus for $0\leq x\leq100$ (which guarantees that $\gamma x/\omega_{0}\ll1$)
and $\omega_{0}t\gg1$, we can approximate
\begin{equation}
Q_{\alpha}(R,t)=j_{\alpha}\left(  x\right)  \int_{-\infty}^{\infty}%
d\omega\,\frac{e^{-i\omega t}-e^{-\gamma t}}{\gamma-i\omega}=\pi j_{\alpha
}\left(  x\right)  e^{-\gamma t}%
\end{equation}
to arrive at the expectation value of the energy density in the WWA,
$\left\langle w^{WW}(R,t)\right\rangle $, given by
\begin{equation}
\left\langle w^{WW}(R,t)\right\rangle =\frac{\hbar\omega_{0}\gamma}{3\pi
c}x^{2}e^{-2\gamma t}\left[  2\left\vert j_{0}\left(  x\right)  \right\vert
^{2}+3\left\vert j_{1}\left(  x\right)  \right\vert ^{2}+\left\vert
j_{2}\left(  x\right)  \right\vert ^{2}\right]  ,\label{wwww}%
\end{equation}
such that
\begin{equation}
\left\langle w^{WW}(R,t)\right\rangle \sim\frac{2\hbar\omega_{0}\gamma}{3\pi
c}x^{2}e^{-2\gamma t}%
\end{equation}
as $R\rightarrow0$. On the other hand, for $x\gg1$, we write the spherical
Bessel function $j_{\alpha}\left[  x\left(  1+\omega/\omega_{0}\right)
\right]  $ in Eq. (\ref{qww}) in terms of spherical Hankel functions and set
$\omega/\omega_{0}=1$ except when it appears in phases to obtain
\begin{equation}
\left\langle w^{WW}(R,t)\right\rangle =\frac{2\hbar\omega_{0}\gamma}%
{c}e^{-2\gamma\left(  t-R/c\right)  }\Theta\left(  t-R/c\right)  ,\label{pass}%
\end{equation}
where $\Theta$ is a Heaviside function, which agrees with Eq. (\ref{wwww}) in
the limit that $x\gg1$ and $\gamma R/c\ll1$ (note that $\gamma R/c=\gamma
x/\omega_{0}$) \cite{pass}. The corresponding result for a classical dipole
may be shown to be \cite{bermcl}%
\begin{equation}
\left\langle w^{class}(R,t)\right\rangle =\frac{2W_{0}\gamma}{c}%
e^{-2\gamma\left(  t-R/c\right)  }\left(  1+\frac{1}{x^{2}}+\frac{3}{2x^{4}%
}\right)  \Theta\left(  t-R/c\right)  ,
\end{equation}
where $W_{0}$ is the original time-averaged energy of the oscillator and the
average is now a time average. The two results agree in the radiation zone,
but not in the near-field region.

\section{Discussion}

The WWA is truly amazing indeed! By taking the spectral density of the vacuum
field to be a constant, including negative frequencies, one arrives at a
theory of spontaneous emission which agrees with experiment, even if the
theory is not quite physically correct. In this paper, we have used the WWA to
calculate the energy and angular momentum in the fields radiated in
spontaneous emission. Both the energy and angular momentum of the fields are
finite at all times, in contrast to the results for a classical point dipole. 

Although we have been comparing the calculation of the total field energy (or
total angular momentum of the field) in spontaneous emission to those of a
radiating point dipole, it is important to point out a fundamental difference
in the two calculations. In calculating the energy radiated by a classical
dipole, one uses expressions involving bilinear products of the field
amplitudes. On the other hand, in the quantum calculation, the average field
energy depends on the expectation value of the product of field operators
rather than on the product of the expectation values of field operators. As
such, strictly speaking, there is no classical analogue of spontaneous emission.

Nevertheless, it appears that two different approaches for calculating the
energy in the field yield two, contradictory results. With the atomic energy,
field energy and interaction energy defined by Eqs. (\ref{hatom})-(\ref{v}),
we are guaranteed that energy is conserved, independent of whether or not the
WWA is used, even if it may be necessary to introduce a cutoff in the sums
over field modes to arrive at finite field and interaction energies. However,
had we used a source-field approach, we would find the total energy in the
field diverges, \textit{even }if we use the RWA and WWA to evaluate the fully
retarded source-field expression for the Poynting vector \cite{pt,stokes} or
energy density \cite{pass,pt}. This dichotomy may be explained by limitations
inherent in a nonrelativistic treatment of spontaneous emission. In
calculating the spectrum of spontaneous emission without the WWA, a cutoff in
the sum over field modes must be used to avoid infinite level shifts.
Moreover, a cutoff in field mode frequencies is consistent with the use of the
dipole approximation. On the other hand, in source-field theory one must sum
over \textit{all} modes of the field to get a fully retarded solution. In the
source-field approach, the field energy is infinite owing to near field
contributions. Thus, the source-field approach (fully -retarded fields,
infinite field energy, no cutoff) and the direct calculation of the
expectation value of the field energy using $\hbar\int d\mathbf{k}%
\sum_{\lambda}\omega_{k}\left\langle a^{\dag}\left(  \mathbf{k}_{\lambda
}\right)  a\left(  \mathbf{k}_{\lambda}\right)  \right\rangle $ (cutoff,
finite field energies, non-fully retarded fields) appear to be incompatible.

There is another problem in using the expressions for the energy and angular
momentum density in dipole approximation to calculate the total energy and
angular momentum in the field. Whereas the calculation of the field energy
density for a point dipole oscillator can be carried out for arbitrary
$R\neq0$, the source-field expressions for the field operators in dipole
approximation are suspect for values of $R$ that are less than atomic
dimensions. As a consequence, integrating the source-field results for the
energy density over all space to get the total field energy may not be
justified. On the other hand, the operator expressions we used for the
\textit{total} energy and angular momentum in the field were derived from the
\textit{exact }field operators, without any approximations. It was only in
calculating the expectation values of these operators that we imposed the
dipole approximation, the RWA and the WWA.

Although the calculation was carried out for atomic emission, the same method
can be used to analyze the problem spontaneous emission by a charged particle
in a constant magnetic field. An outline of the solution is given in Appendix B.

The research of PRB and AK is supported by the National Science Foundation and
the Air Force Office of Scientific Research.

\bigskip

\appendix

\section{Field Angular Momentum}

In this appendix we express the angular momentum of the field in terms of
creation and annihilation operators. We start with the expression for the
Poynting vector,%
\begin{align}
\mathbf{S}(\mathbf{R}) &  =\epsilon_{0}c^{2}\mathbf{\,E}(\mathbf{R}%
)\times\mathbf{B}(\mathbf{R})\nonumber\\
&  =-\frac{c}{\left(  2\pi\right)  ^{3}}\left(  \frac{\hbar}{2}\right)  \int
d\mathbf{k\,}\sqrt{\omega_{k}}\sum_{\lambda}\left[  a\left(  \mathbf{k}%
_{\lambda}\right)  e^{i\mathbf{k}\cdot\mathbf{R}}-a^{\dag}\left(
\mathbf{k}_{\lambda}\right)  e^{-i\mathbf{k}\cdot\mathbf{R}}\right]
\boldsymbol{\epsilon}_{\mathbf{k}}^{(\lambda)}\nonumber\\
&  \times\int d\mathbf{k}^{\prime}\mathbf{\,}\sqrt{\omega_{k^{\prime}}}%
\sum_{\lambda^{\prime}}\left[  a\left(  \mathbf{k}_{\lambda^{\prime}}^{\prime
}\right)  e^{i\mathbf{k}^{\prime}\cdot\mathbf{R}}-a^{\dag}\left(
\mathbf{k}_{\lambda^{\prime}}^{\prime}\right)  e^{-i\mathbf{k}^{\prime}%
\cdot\mathbf{R}}\right]  \left(  \mathbf{\hat{k}}^{\prime}\mathbf{\times
}\boldsymbol{\epsilon}_{\mathbf{k}^{\prime}}^{(\lambda^{\prime})}\right)
.\label{a1}%
\end{align}
Using Eq. (\ref{comm}) to reorder the term proportional to $a\left(
\mathbf{k}_{\lambda}\right)  a^{\dag}\left(  \mathbf{k}_{\lambda^{\prime}%
}^{\prime}\right)  $ in Eq. (\ref{a1}), we find that the delta function
contribution from Eq. (\ref{comm}) varies as%
\begin{align}
&  -\frac{c}{\left(  2\pi\right)  ^{3}}\left(  \frac{\hbar}{2}\right)
\sum_{\mathbf{\lambda},\lambda^{\prime}}\int d\mathbf{k}\int d\mathbf{k}%
^{\prime}\sqrt{\omega_{k}\omega_{k^{\prime}}}\nonumber\\
&  \times e^{i\mathbf{k}\cdot\mathbf{R}}e^{-i\mathbf{k}^{\prime}%
\cdot\mathbf{R}}\boldsymbol{\epsilon}_{\mathbf{k}}^{(\lambda)}\times\left(
\mathbf{\hat{k}}^{\prime}\mathbf{\times}\boldsymbol{\epsilon}_{\mathbf{k}%
^{\prime}}^{(\lambda^{\prime})}\right)  \delta\left(  \mathbf{k}%
-\mathbf{k}^{\prime}\right)  \delta_{\lambda,\lambda^{\prime}}\nonumber\\
&  =-\frac{c}{\left(  2\pi\right)  ^{3}}\left(  \frac{\hbar}{2}\right)
\sum_{\mathbf{\lambda},\lambda^{\prime}}\int d\mathbf{k\,}\omega
_{k}\boldsymbol{\epsilon}_{\mathbf{k}}^{(\lambda)}\times\left(  \mathbf{\hat
{k}\times}\boldsymbol{\epsilon}_{\mathbf{k}}^{(\lambda)}\right)  =0,
\end{align}
since the angular integral vanishes. We are left with%
\begin{align}
\mathbf{S}(\mathbf{R}) &  =-\frac{c}{\left(  2\pi\right)  ^{3}}\left(
\frac{\hbar}{2}\right)  \int d\mathbf{k}\int d\mathbf{k}^{\prime}\sum
_{\lambda,\lambda^{\prime}}\boldsymbol{\epsilon}_{\mathbf{k}}^{(\lambda
)}\times\left(  \mathbf{\hat{k}}^{\prime}\mathbf{\times}\boldsymbol{\epsilon
}_{\mathbf{k}^{\prime}}^{(\lambda^{\prime})}\right)  \sqrt{\omega_{k}%
\omega_{k^{\prime}}}\nonumber\\
&  \times\left(
\begin{array}
[c]{c}%
a\left(  \mathbf{k}_{\lambda}\right)  a\left(  \mathbf{k}_{\lambda^{\prime}%
}^{\prime}\right)  e^{i\mathbf{k}\cdot\mathbf{R}}e^{i\mathbf{k}^{\prime}%
\cdot\mathbf{R}}-a^{\dag}\left(  \mathbf{k}_{\lambda^{\prime}}^{\prime
}\right)  a\left(  \mathbf{k}_{\lambda}\right)  e^{-i\mathbf{k}^{\prime}%
\cdot\mathbf{R}}e^{i\mathbf{k}\cdot\mathbf{R}}\\
-a^{\dag}\left(  \mathbf{k}_{\lambda}\right)  a\left(  \mathbf{k}%
_{\lambda^{\prime}}^{\prime}\right)  e^{-i\mathbf{k}\cdot\mathbf{R}%
}e^{i\mathbf{k}^{\prime}\cdot\mathbf{R}}+a^{\dag}\left(  \mathbf{k}_{\lambda
}\right)  a^{\dag}\left(  \mathbf{k}_{\lambda^{\prime}}^{\prime}\right)
e^{-i\mathbf{k}\cdot\mathbf{R}}e^{-i\mathbf{k}^{\prime}\cdot\mathbf{R}}%
\end{array}
\right)  .\label{poynt1}%
\end{align}
Now each term is normal ordered and we can neglect terms varying as
$a_{\mathbf{k}_{\lambda}}a_{\mathbf{k}_{\lambda^{\prime}}^{\prime}}$ or
$a_{\mathbf{k}_{\lambda}}^{\dag}a_{\mathbf{k}_{\lambda^{\prime}}^{\prime}%
}^{\dag}$ since they will vanish when taking expectation values. Combining
Eqs. (\ref{eltot}) and (\ref{poynt1}), we find%
\begin{align}
\mathbf{L}_{f} &  =\hbar c\frac{\omega_{0}}{2\left(  2\pi\right)  ^{3}}\int
d\mathbf{k}\int d\mathbf{k}^{\prime}\sum_{\lambda,\lambda^{\prime}}a^{\dag
}\left(  \mathbf{k}_{\lambda^{\prime}}^{\prime}\right)  a\left(
\mathbf{k}_{\lambda}\right)  \sqrt{\omega_{k}\omega_{k^{\prime}}}\nonumber\\
&  \times\left[  \boldsymbol{\epsilon}_{\mathbf{k}}^{(\lambda)}\times\left(
\mathbf{\hat{k}}^{\prime}\mathbf{\times}\boldsymbol{\epsilon}_{\mathbf{k}%
^{\prime}}^{(\lambda^{\prime})}\right)  \right]  \times\int d\mathbf{R\,R}%
e^{i\left(  \mathbf{k}-\mathbf{k}^{\prime}\right)  \cdot\mathbf{R}%
}+adj.\label{poy}%
\end{align}
But
\begin{equation}
\int d\mathbf{R\,R}e^{i\left(  \mathbf{k}-\mathbf{k}^{\prime}\right)
\cdot\mathbf{R}}=-i\left(  2\pi\right)  ^{3}\mathbf{\nabla}_{\mathbf{k}}%
\delta\left(  \mathbf{k}-\mathbf{k}^{\prime}\right)  ,
\end{equation}
such that%
\begin{align}
\mathbf{L}_{f} &  =\frac{i\hbar c}{2}\sum_{\lambda,\lambda^{\prime}}\int
d\mathbf{k}\int d\mathbf{k}^{\prime}\sqrt{\omega_{k}\omega_{k^{\prime}}%
}\nonumber\\
&  \times a^{\dag}\left(  \mathbf{k}_{\lambda^{\prime}}^{\prime}\right)
a\left(  \mathbf{k}_{\lambda}\right)  \left[  \boldsymbol{\epsilon
}_{\mathbf{k}}^{(\lambda)}\times\left(  \mathbf{\hat{k}}^{\prime
}\mathbf{\times}\boldsymbol{\epsilon}_{\mathbf{k}^{\prime}}^{(\lambda^{\prime
})}\right)  \right]  \times\mathbf{\nabla}_{\mathbf{k}}\delta\left(
\mathbf{k}-\mathbf{k}^{\prime}\right)  +adj\nonumber\\
&  =i\frac{\hbar}{2c}\sum_{\lambda,\lambda^{\prime}}\int d\mathbf{k\int
d\mathbf{k}^{\prime}}\delta\left(  \mathbf{k}-\mathbf{k}^{\prime}\right)
\mathbf{\nabla}_{\mathbf{k}}\times\left\{  \sqrt{\omega_{k}\omega_{k^{\prime}%
}}a^{\dag}\left(  \mathbf{k}_{\lambda^{\prime}}^{\prime}\right)  a\left(
\mathbf{k}_{\lambda}\right)  \left[  \boldsymbol{\epsilon}_{\mathbf{k}%
}^{(\lambda)}\times\left(  \mathbf{\hat{k}}^{\prime}\mathbf{\times
}\boldsymbol{\epsilon}_{\mathbf{k}^{\prime}}^{(\lambda^{\prime})}\right)
\right]  \right\}  +adj\nonumber\\
&  =\left(  i\frac{\hbar}{2c}\sum_{\lambda,\lambda^{\prime}}\int
d\mathbf{k\,}\sqrt{\omega_{k}}a^{\dag}\mathbf{\left(  \mathbf{k}%
_{\lambda^{\prime}}\right)  }\left[  \left(  \mathbf{\hat{k}\times
}\boldsymbol{\epsilon}_{\mathbf{k}}^{(\lambda^{\prime})}\right)
\cdot\mathbf{\nabla}_{\mathbf{k}}\right]  \left[  \sqrt{\omega_{k}}a\left(
\mathbf{k}_{\lambda}\right)  \boldsymbol{\epsilon}_{\mathbf{k}}^{(\lambda
)}\right]  +adj\right)  \nonumber\\
&  +\left(  -i\frac{\hbar}{2c}\sum_{\lambda,\lambda^{\prime}}\int
d\mathbf{k\,}\sqrt{\omega_{k}}a^{\dag}\mathbf{\left(  \mathbf{k}%
_{\lambda^{\prime}}\right)  }\left(  \mathbf{\hat{k}\times}%
\boldsymbol{\epsilon}_{\mathbf{k}}^{(\lambda^{\prime})}\right)  \mathbf{\nabla
}_{\mathbf{k}}\cdot\left[  \sqrt{\omega_{k}}a\left(  \mathbf{k}_{\lambda
}\right)  \boldsymbol{\epsilon}_{\mathbf{k}}^{(\lambda)}\right]  +adj\right)
,\label{elefap1}%
\end{align}
where we have used the identity $\mathbf{F}\times\mathbf{\nabla}%
g=-\mathbf{\nabla}\times\left(  g\mathbf{F}\right)  +g\mathbf{\nabla}%
\times\mathbf{F}$ and the fact that the $\mathbf{\nabla}\times\left(
g\mathbf{F}\right)  $ term vanishes when converted to a surface integral.

\section{WWA for Cyclotron Motion}

We now consider a particle having mass $M$ and charge $q$ undergoing
two-dimensional motion in the $x-y$ plane in the presence of a magnetic field
$\mathbf{B}=B\mathbf{\hat{z}}$, where $B>0$ and a caret over a character
indicates a unit vector. The Hamiltonian in this case is still given by Eq.
(\ref{ham1}), with $H_{atom}$ replaced by
\begin{equation}
H_{q}=-\frac{\hbar^{2}}{2M\rho}\frac{\partial}{\partial\rho}\left(  \rho
\frac{\partial}{\partial\rho}\right)  +\frac{L_{z}^{2}}{2M\rho^{2}}%
+\frac{q^{2}B^{2}\rho^{2}}{8M}-\frac{qBL_{z}}{2M},
\end{equation}
where $L_{z}=-i\hbar\partial/\partial\phi$ is an angular momentum operator and
$\rho$ and $\phi$ are cylindrical coordinate. The ground state wave function
having energy $\hbar\omega_{c}/2$ is \cite{land}%
\begin{equation}
\psi_{g}(\rho)=\frac{e^{-\rho^{2}/(2a^{2})}}{\sqrt{\pi}a}\label{gg}%
\end{equation}
and the first excited state having energy $3\hbar\omega_{c}/2$ and angular
momentum $\hbar$ is
\begin{equation}
\psi_{e}(\rho)=\rho\frac{e^{-\rho^{2}/(2a^{2})}}{a^{2}\sqrt{\pi}}e^{i\phi
},\label{ee}%
\end{equation}
where%
\begin{equation}
\omega_{c}=\frac{qB}{M}%
\end{equation}
is the cyclotron frequency and%
\begin{equation}
a=\sqrt{\frac{2\hbar}{M\omega_{c}}}=\sqrt{\frac{2\hbar}{qB}}.
\end{equation}
Limiting our calculation to the two states with wave functions given by Eqs.
(\ref{gg}) and (\ref{ee}), we can write the interaction Hamiltonian (\ref{v})
in dipole approximation as
\begin{equation}
V=-\frac{i}{\left(  2\pi\right)  ^{3/2}}\int d\mathbf{k}\left(  \frac
{\hbar\omega_{k}}{2\epsilon_{0}}\right)  ^{1/2}\sum_{\lambda}\left[
\boldsymbol{\mu}_{eg}\cdot\boldsymbol{\epsilon}_{\mathbf{k}}^{(\lambda
)}a\left(  \mathbf{k}_{\lambda}\right)  -a^{\dag}\left(  \mathbf{k}_{\lambda
}\right)  \boldsymbol{\mu}_{ge}\cdot\boldsymbol{\epsilon}_{\mathbf{k}%
}^{(\lambda)}\right]  ,
\end{equation}
where
\begin{subequations}
\begin{align}
\boldsymbol{\mu}_{ge}\cdot\boldsymbol{\epsilon}_{\mathbf{k}}^{(\theta)} &
=\left[  \boldsymbol{\mu}_{eg}\cdot\boldsymbol{\epsilon}_{\mathbf{k}}%
^{(\theta)}\right]  ^{\ast}=\frac{a}{2}\cos\theta_{k}e^{i\phi_{k}}\\
\boldsymbol{\mu}_{ge}\cdot\boldsymbol{\epsilon}_{\mathbf{k}}^{(\phi)} &
=\left[  \boldsymbol{\mu}_{eg}\cdot\boldsymbol{\epsilon}_{\mathbf{k}}^{(\phi
)}\right]  ^{\ast}=i\frac{a}{2}e^{i\phi_{k}}%
\end{align}
and $\boldsymbol{\mu}_{ge}$ and $\boldsymbol{\mu}_{ge}$ are dipole matrix elements.

The calculation of spontaneous emission from the excited to ground state then
proceeds as in the atomic case with
\end{subequations}
\begin{equation}
c_{e}(t)=e^{-\gamma_{c}t},
\end{equation}%
\begin{equation}
\gamma_{c}=\frac{q^{2}\omega_{c}^{2}}{6\pi\epsilon_{0}Mc^{3}},
\end{equation}
and
\begin{align}
c_{g}^{(\theta)}(\mathbf{k},t) &  =\frac{i}{\left(  2\pi\right)  ^{3/2}%
}\left(  \frac{\omega_{k}}{2\hbar\epsilon_{0}}\right)  ^{1/2}\left(  \frac
{a}{2}\cos\theta_{k}e^{i\phi_{k}}\right)  e^{-i\left(  \omega_{c}-\omega
_{k}\right)  t}\nonumber\\
&  \times\frac{e^{-i\left(  \omega_{k}-\omega_{0}\right)  t}-e^{-\gamma t}%
}{\gamma-i\left(  \omega_{k}-\omega_{0}\right)  },\\
c_{g}^{(\phi)}(\mathbf{k},t) &  =\frac{i}{\left(  2\pi\right)  ^{3/2}}\left(
\frac{\omega_{k}}{2\hbar\epsilon_{0}}\right)  ^{1/2}\left(  i\frac{a}%
{2}e^{i\phi_{k}}\right)  e^{-i\left(  \omega_{c}-\omega_{k}\right)
t}\nonumber\\
&  \times\frac{e^{-i\left(  \omega_{k}-\omega_{0}\right)  t}-e^{-\gamma t}%
}{\gamma-i\left(  \omega_{k}-\omega_{0}\right)  }.
\end{align}
It then follows that, in the WWA, the average energy of the charge and the
average energy of the transverse radiation field are given by
\begin{subequations}
\begin{align}
\left\langle H_{q}(t)\right\rangle  &  =\hbar\omega_{c}\left\vert
c_{e}(t)\right\vert ^{2}=\hbar\omega_{c}e^{-2\gamma_{c}t},\\
\left\langle H_{f}(t)\right\rangle  &  =\hbar\int d\mathbf{k}\omega_{k}%
\sum_{\lambda}\left\vert c_{m_{G}}^{(\lambda)}(\mathbf{k},t)\right\vert
^{2}=\hbar\omega_{c}\left(  1-e^{-2\gamma_{c}t}\right)  ,
\end{align}
and the expectation value of the total angular momentum in the field,
calculated using Eq. (\ref{angtot}), is
\end{subequations}
\begin{align}
\left\langle L_{f}(t)\right\rangle _{z} &  =-i\frac{\hbar}{2}\int
d\mathbf{k\,}c_{g}^{(\phi)}(\mathbf{k},t)^{\ast}\left[  \frac{\partial
}{\partial\theta_{k}}\left(  \sin\theta_{k}c_{g}^{(\theta)}(\mathbf{k}%
,t)\right)  +\frac{\partial}{\partial\phi_{k}}\left(  c_{g}^{(\phi
)}(\mathbf{k},t)\right)  \right]  +c.c.\nonumber\\
&  =-i\frac{\hbar}{2}\int d\mathbf{k\,}c_{g}^{(\phi)}(\mathbf{k},t)^{\ast
}\left[  c_{g}^{(\theta)}(\mathbf{k},t)\left(  \cos^{2}\theta_{k}-\sin
^{2}\theta_{k}\right)  -ic_{g}^{(\phi)}(\mathbf{k},t)\right]  +c.c.\nonumber\\
&  =\hbar\left(  1-e^{-2\gamma_{c}t}\right)  .
\end{align}
Energy and angular momentum are conserved for the charge-field system, in
contrast to a classical calculation in which only \textit{half} of the
charge's initial angular momentum is transferred to the field \cite{bermcl}.

\bigskip

\begin{thebibliography}{99}                                                                                               %


\bibitem {ww}Weisskopf, V., \& Wigner, E. (1930). Berechnung der nat\"{u}auf
grund der diracschen lichttheorie (Calculation of the natural line width on
the basis of Dirac's theory of light). Zeitschrift fur Physik, 92, 54-73. This
article is translated by J. B. Sykes and reprinted in Hindmarsh, W. (1967).
Atomic Spectra (pp. 304-327). Oxford: Pergamon Press.

\bibitem {bermcl}P. R. Berman, A. Kuzmich, and P. W. Milonni, Classical dipole
radiation - revisited, Phys. Rev. A\textbf{111}, 013528 (2025).

\bibitem {frbar}S. Franke and S. M. Barnett, Angular momentum in spontaneous
emission, J. Phys. B \textbf{29}, 2141 (1996).

\bibitem {andr}D. L. Andrews, Optical angular momentum: Multipole transitions
and photonics, Phys. Rev. A \textbf{81}, 033825 (2010).

\bibitem {barn2}S. M. Barnett, F. C. Speirits, and M. Babiker, Optical angular
momentum in atomic transitions: a paradox, J. Phys. A \textbf{55}, 234008 (2022).

\bibitem {bf}P. R. Berman and G. H. W. Ford, Spontaneous Decay, Unitarity, and
the Weisskopf-Wigner Approximation, Advances in Atomic, Molecular, and Optical
Physics \textbf{59}, 175 (2010).

\bibitem {simm}J. W. Simmons and M. J. Guttmann, \textit{States, Waves, and
Photons: A Modern Introduction to Light} (Addison-Wesley, Reading, MA, 1970)
pp. 266-270.

\bibitem {mc3}K. T. McDonald, Orbital and Spin Angular Momentum of
Electromagnetic Fields, http://kirkmcd.princeton.edu/examples/spin.pdf (2021).

\bibitem {cc}C. Cohen-Tannoudji, J. Dupont-Roc, and G. Gilbert,
\textit{Photons and Atoms: Introduction to Quantum Electrodynamics }(John
Wiley and Sons, Strauss GmbH, Morlenbach,1989) pp. 47-49.

\bibitem {enk}S. J. van Enk and G. Nienhuis, Spin and Orbital Angular Momentum
of Photons, Europhys. Lett. \textbf{25}, 497 (1994).

\bibitem {bw}L. Mandel and E. Wolf, Optical coherence and quantum optics
(Cambridge University Press, New York, 1995) pp. 485-491.

\bibitem {barn3}S. M. Barnett, L. Allen, R. P. Cameron, C. R. Gilson, M. J.
Padgett, F. C. Speirits, and A. M. Yao, On the nature of the spin and orbital
angular momentum, J. Opt. \textbf{18}, 064004 (2016).

\bibitem {mil}P. W. Milonni, D. P. V. James, and H. Fearn, Photodetection and
causality in quantum optics, Phys. Rev. A \textbf{52}, 1525 (1995), and
references therein

\bibitem {barn}S. M. Barnett, Optical angular-momentum flux, J. Opt. B:
Quantum Semiclass. Opt. \textbf{4}, S7 (2002).

\bibitem {berww}P. R. Berman, Wigner-Weisskopf approximation under typical
experimental conditions, Phys. Rev. A \textbf{72}, 025804 (2005).

\bibitem {pt}E. A. Power and T. Thirunamachandran, Quantum electrodynamics
with nonrelativistic sources. IV. Poynting vector, energy densities, and other
quadratic operators of the electromagnetic field, Phys. Rev. A \textbf{45}, 54 (1992).

\bibitem {stokes}A. Stokes, Vacuum source-field correlations and advanced
waves in quantum optics, Quantum \textbf{2}, 46 (2018). Stokes starts from the
source-field expression for the fields, but evaluates interference terms
involving the free and source fields using the RWA and WWA. His calculation is
restricted to the far field, but his approach would yield Poynting vectors
that diverge as $R\rightarrow0$.

\bibitem {pass}G. Compagno, R. Passante, and F. Persico, Virtual photons,
causality, and atomic dynamics in spontaneous emission, J. Mod. Optics
\textbf{37:8}, 1377 (2007). In this paper the authors obtain an expression for
the energy density in spontaneous emission using perturbation theory without
using the RWA for an initial state of the atom excited and no photons in the
field. . Their result was fully-retarded and agrees with Eq. (\ref{pass}) when
$x\gg1$ and $\gamma R/c\ll1$. They did not evaluate their expression for
$x\ll1$, but it appears to be divergent as $R\rightarrow0$.

\bibitem {land}O. Ciftja, Detailed solution of the problem of Landau states in
a symmetric gauge, Eur. J. Phys. \textbf{41}, 035404 (2020).
\end{thebibliography}
\end{document}